\documentclass[lettersize,journal]{IEEEtran}
\usepackage{amsmath,amsfonts}
\usepackage{algorithmic}
\usepackage{algorithm}
\usepackage{array}
\usepackage[caption=false,font=normalsize,labelfont=sf,textfont=sf]{subfig}
\usepackage{textcomp}
\usepackage{stfloats}
\usepackage{url}
\usepackage{verbatim}
\usepackage{graphicx}
\usepackage{cite}
\usepackage{subfig}
\usepackage{bm}
\usepackage{booktabs}
\usepackage{multirow}
\begin{document}

\title{ISMAF: Intrinsic-Social Modality Alignment and Fusion for Multimodal Rumor Detection}

\author{Zihao Yu, Xiang Li, Jing~Zhang$^\ast$
	\IEEEcompsocitemizethanks{Z. Yu, X. Li, and J. Zhang are with the School of Cyber Science and Engineering, Southeast University, No. 2 SEU Road, Nanjing 211189, China. (Email: \{220235336,  lixiang\_15, jingz\}@seu.edu.cn)}
	\thanks{Dr. Jing Zhang is the corresponding author.}
}

\markboth{Journal of \LaTeX\ Class Files,~Vol.~99, No.~9, May~2025}%
{Zhang \MakeLowercase{\textit{et al.}}: A Sample Article Using IEEEtran.cls for IEEE Journals}

\IEEEpubid{0000--0000/00\$00.00~\copyright~2025 IEEE}

\maketitle

\begin{abstract}
 The rapid dissemination of rumors on social media highlights the urgent need for automatic detection methods to safeguard societal trust and stability. While existing multimodal rumor detection models primarily emphasize capturing consistency between intrinsic modalities (e.g., news text and images), they often overlook the intricate interplay between intrinsic and social modalities. This limitation hampers the ability to fully capture nuanced relationships that are crucial for comprehensive understanding. Additionally, current methods struggle with effectively fusing social context with textual and visual information, resulting in fragmented interpretations. To address these challenges, this paper proposes a novel Intrinsic-Social Modality Alignment and Fusion (ISMAF) framework for multimodal rumor detection. ISMAF first employs a cross-modal consistency alignment strategy to align complex interactions between intrinsic and social modalities. It then leverages a mutual learning approach to facilitate collaborative refinement and integration of complementary information across modalities. Finally, an adaptive fusion mechanism is incorporated to dynamically adjust the contribution of each modality, tackling the complexities of three-modality fusion. Extensive experiments on both English and Chinese real-world multimedia datasets demonstrate that ISMAF consistently outperforms state-of-the-art models.
\end{abstract}


\section{Introduction}
\IEEEPARstart{W}{ith} the rise of the digital era, social media platforms have revolutionized the dissemination of information, facilitating global connectivity and broadening access to knowledge. However, the revolution comes with a cost: social media platforms have become a significant source of misinformation dissemination\cite{sharma2019combating, shu2017fake}, which can mislead the public, cause economic harm, and even incite unrest. According to a study\cite{ngadiron2020spread} in 2019, the global economic loss attributed to misinformation reached a staggering 78 billion dollars annually. Thus, the ability to accurately detect misinformation is crucial for maintaining societal trust and well-being.

Traditional misinformation detection methods have primarily relied on unimodal textual analysis, utilizing techniques such as Recurrent Neural Networks (RNNs) and Convolutional Neural Networks (CNNs) to classify content based solely on linguistic patterns\cite{ma2016detecting, yu2017convolutional}. Although these text-based approaches have achieved some level of success, they often overlook the valuable information present in other modalities, particularly visual content. This limitation has prompted a shift towards multimodal detection methods\cite{wang2018eann, khattar2019mvae, singhal2019spotfake, zhou2020safe}, which integrate both textual and visual features to capture a more comprehensive understanding of the information. Recent studies\cite{jin2016novel, qi2019exploiting} have demonstrated that such multimodal approaches can significantly improve detection accuracy by leveraging the complementary strengths of different data sources.

\IEEEpubidadjcol

On the other hand, the exploitation of social context information has revealed rich contextual dimensions that contribute to rumor detection. Many studies have focused specifically on social context information, considering its various aspects such as time series\cite{ma2015detect}, user reactions\cite{yang2021rumor, dou2021user}, social event interactions\cite{he2021rumor} and propagation patterns\cite{liu2018early, guo2018rumor, yuan2019jointly, fang2023unsupervised}. Given the significant auxiliary effectiveness of social context information in the above methods, recent studies\cite{jin2017multimodal, zheng2022mfan, xu2024clffrd} have considered integrating social context features with textual and visual features to further enhance detection performance. Previous work \cite{ zheng2022mfan, xu2024clffrd } commonly use social context information to complement the news contents (texts and images) by alignment and data fusion strategies. 

\begin{figure}[t]  
	\centering  
	\captionsetup{font=small}
	\includegraphics[width=0.7\columnwidth]{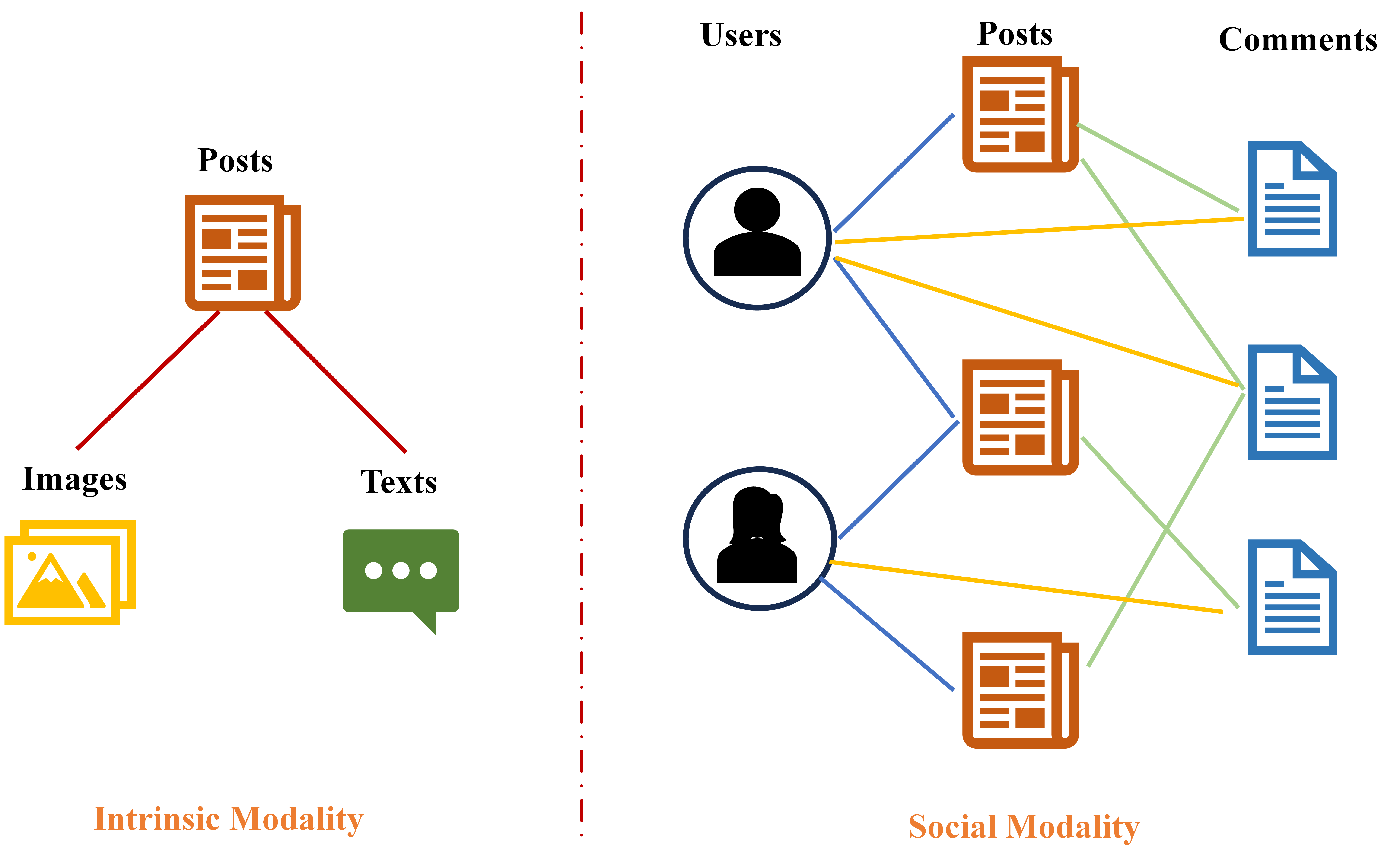} 
	\caption{Intrinsic modality and social modality of multimodal posts.}  
	\label{introduction}
	\vspace{-15pt}
\end{figure}

Despite advancements, essentially, they ignore the \textit{intrinsic-social inconsistency}. For instance, in the real world, individuals’ judgments about the veracity of news are influenced by two primary modalities: intrinsic and social. As shown in Fig.~\ref{introduction}, \textit{Intrinsic modality} encompasses the core content of news, such as text and images, while \textit{social modality} includes social signals like propagation patterns and user interactions. However, in many cases, these two sources of information may present inconsistent cues. A piece of news that appears credible based on its well-written text or convincing images may originate from a source known for propagating misinformation, casting doubt on its credibility. Conversely, dubious content may gain legitimacy through circulation among reputable users. Similarly, the inconsistency between intrinsic modality and social modality creates ambiguity, which traditional methods are inadequately equipped to address.

This study aims to capture the deep interactions and correspondence between the intrinsic modality and social modality of news content, enabling the learning of comprehensive multimodal representations. This issue is both critical and challenging due to the following factors:

1) \textit{Limited Exploration of Intrinsic-Social Correspondence}: While existing studies\cite{qian2021hierarchical, xu2024hierarchical} have concentrated on the consistency between news text and images, the exploration of the correspondence between news intrinsic modality and social modality remains relatively unexplored. Both intrinsic and social modalities provide complementary insights, offering a more holistic understanding of news content.

2) \textit{Challenges in Three-Modality Fusion}: Although substantial progress has been made in fusing textual and visual modalities through techniques such as concatenation operation\cite{khattar2019mvae, singhal2022leveraging} and attention-based methods\cite{wu2021multimodal, zheng2022mfan}, these approaches often struggle to capture the complex interdependencies among the three modalities due to their reliance on fixed fusion patterns. The increased complexity of multimodal news content introduces further challenges in achieving effective fusion across all modalities, highlighting the need for more sophisticated and dynamic fusion strategies.

To address the above challenges, we propose a concise yet efficient Intrinsic-Social Modality Alignment and Fusion (ISMAF) framework for multimodal rumor detection. For each post within rumor datasets, we extract the intrinsic modality features from its text and image, and the social modality features from the constructed social graph. To specifically address the challenge of limited exploration of intrinsic-social correspondence, we design a unified scheme for bridging these modalities through two complementary modules. The first module employs a cross-modal consistency alignment strategy to align the complex interactions between intrinsic and social modalities. The second one leverages a mutual learning approach to facilitate collaborative learning and refinement between these modalities, enabling the effective integration of their complementary information. Additionally, to address the complexities of three-modality fusion, we incorporate an adaptive fusion mechanism that dynamically adjusts the contribution of each modality based on its relevance to the input. This design enables the generation of flexible and robust multimodal representations, thereby enhancing the overall performance. The main contributions of this study are as follows:
\begin{itemize} 
	\item ISMAF is the first rumor detection framework to explicitly explore the correspondence between intrinsic and social modalities in news content. By employing a cross-modal consistency alignment strategy and a mutual learning approach, ISMAF encourages the intrinsic features to be embedded closely with their corresponding social features, thereby enhancing the model's ability to capture complex interdependencies and improving overall detection performance.
	\item We introduce an adaptive fusion mechanism that dynamically adjusts the contribution of each modality according to its informativeness, allowing the framework to capture subtle cross-modal interactions and generate more expressive multimodal representations.
	\item Extensive experiments on two real-world datasets consistently demonstrate that our ISMAF framework outperforms state-of-the-art baselines in rumor detection tasks, highlighting its effectiveness in capturing the complex interplay between intrinsic and social modalities.
\end{itemize}

The remainder of the paper is organized as follows. Section~\ref{sec:rw} briefly reviews existing rumor detection approaches. Section~\ref{sec:ps} formally defines the research problem. Section~\ref{sec:mtd} presents the proposed ISMAF framework in detail. Section~\ref{sec:exp} presents the experimental results and discussion. Section~\ref{sec:con} concludes the paper and outlines future research directions.

\section{Related Work}\label{sec:rw}
\subsection{Unimodal Rumor Detection}
Unimodal rumor detection methods focus on extracting features solely from the news text content. These methods can be broadly categorized into content-based approaches and social context-based approaches, both of which utilize information derived from the textual content in different ways.

\textit{Content-based approaches} center on the analysis of the linguistic and stylistic features of news content \cite{perez2017automatic}. Several studies explored detecting fake news by identifying distinct writing styles, such as lexical, syntactic, and semantic features, between real and fake news \cite{hualing2023survey}. For example, Rubin et al.\cite{rubin2015towards} demonstrated that certain punctuation patterns can effectively distinguish deceptive texts from truthful ones. Potthast et al. \cite{potthast2017stylometric} emphasized the importance of features such as character unigrams and readability scores in identifying fake news. Similarly, Perez et al. \cite{perez2017automatic} employed machine learning models to leverage psycholinguistic and syntactic features.

\textit{Social context-based approaches} extend beyond direct textual analysis by integrating information related to user interactions and engagement with the news content. Some of these approaches incorporate auxiliary textual elements derived from user-generated content, such as comments \cite{shu2019defend, rao2021stanker}, assessments of user credibility \cite{mukherjee2015leveraging, yuan2020early}, and emotional expressions \cite{xue2022multi, zhang2021mining}. Concretely, Shu et al. \cite{shu2019defend} combined news content with user comments within an explainable framework, while Li et al. \cite{li2019rumor} incorporated user credibility information. Zhang et al. \cite{zhang2021mining} explored the relationship between publisher emotions and social emotions to refine detection performance.

Others focus on capturing the propagation patterns and structural relationships of news on social media platforms. Recognizing the significance of the news environment\cite{sheng2022zoom} information, these approaches aim to extract propagation-based features that characterize how information spreads.  For instance,  Ma\cite{ma2018rumor} employed tree-structured recursive neural networks to integrate both propagation structure and content semantics. Bian et al.\cite{bian2020rumor} developed a bi-directional graph convolutional networks to capture structural relations from news propagation graphs. Wei et al.\cite{wei2021towards} propose a graph-based method to handle the uncertainty issue in propagation.

\subsection{Multimodal Rumor Detection}
While unimodal methods can be effective in certain scenarios, they are often limited by their reliance solely on textual features, failing to fully exploit the potential of visual and other multimodal information. Prior studies\cite{alam2021survey, chen2020hgmf} have highlighted the benefits of multimodal approaches, demonstrating that combining information from multiple modalities can significantly enhance detection performance.

Recent advance in multimodal approaches focused on leveraging both textual and visual features to improve rumor detection. For example, Khattar et al. \cite{khattar2019mvae} proposed a fake news detection framework based on a variational auto-encoder that leverages multimodal representations through a specialized decoder. Singhal et al. \cite{singhal2019spotfake} applied pre-trained BERT to extract text features and VGG model to extract image features, subsequently concatenating these modalities for classification. Zhou et al.\cite{zhou2020safe} introduced a novel method that utilizes cosine similarity between text and image modalities to capture intermodal relationships. Addressing the challenge of mismatched image-text pairs, Xue et al.\cite{xue2021detecting} emphasized the importance of consistency between textual and visual information. Furthermore, Dhawan et al. \cite{dhawan2024game} employed a graph attention-based framework to model interactions between textual and visual features, while Guo et al. \cite{guo2024rumor} developed a review-based fusion mechanism to integrate these features effectively.

Beyond the fundamental integration of textual and visual features, several studies have incorporated additional social context information to enhance detection capabilities. Jin et al. \cite{jin2017multimodal} extracted hashtags, user interactions, and sentiment from social media platforms to construct an initial social context representation. They then employed a recurrent neural network with an attention mechanism to fuse this social context with text and image features from social media posts. Zheng et al. \cite{zheng2022mfan} made a significant advancement by employing a graph-based model that constructs a social graph in which nodes represent users, posts, and comments, facilitating the extraction of social context features. Xu et al.\cite{xu2024clffrd} further refined this graph-based approach by introducing a curriculum learning strategy to automate sample selection and training, thereby enhancing the model's overall performance.

These multimodal approaches significantly improve over unimodal approaches by effectively integrating diverse features of news content\cite{wu2023human}. However, despite these improvements, existing methods exhibit several limitations. Many of them\cite{zhou2020safe,xue2021detecting, xu2024hierarchical} primarily focus on the consistency within intrinsic modalities, such as news text and images, while neglecting the intricate correspondence with social modalities. This oversight restricts the ability to capture nuanced relationships that could enhance overall understanding. On the other hand, methods\cite{jin2017multimodal,zheng2022mfan} that incorporate social modalities often struggle to effectively fuse social context with textual and visual information, which can result in a fragmented comprehension of the presented information. To address these issues, we employ a unified scheme to bridge modalities and an adaptive fusion strategy within the ISMAF framework in this paper.

\section{Problem Statement}\label{sec:ps}
Let \( P = \{p_1, p_2, \dots, p_M\} \) represent a set of multimedia posts on social media, where each post \( p_i \in P \) consists of the following elements: \( t_i \), \( v_i \), \( u_i \), and \( c_i \). Specifically, \( t_i \) refers to the text content, \( v_i \) denotes the associated image, \( u_i \) represents the user who posted \( p_i \), and \( c_i = \{c_{i1}, c_{i2}, \dots, c_{ij}\} \) represents a set of comments related to the post. Each comment \( c_{ij} \) is made by a corresponding user \( u_{ij} \).

The goal of the rumor detection task is to predict whether a post \( p_i \) is a rumor or not. It is formulated as a binary classification problem, where \( y \in \{0, 1\} \) is the label assigned to each post. Specifically, \( y = 1 \) indicates that the post is a rumor, and \( y = 0 \) indicates a non-rumor. The objective is to learn a function \( f(p_i) \rightarrow y \) that maps each post \( p_i \) to its corresponding label.

Table \ref{tab:notations} summarizes the notations appearing in this paper.

\begin{table}[htbp]
	\caption{Notations and Descriptions}
	\centering
	\begin{tabular}{r|p{0.7\linewidth}} 
		\hline
		\textbf{Notation} & \textbf{Description} \\
		\hline
		$ R_T^i $ & textual feature vector of post $p_i$ \\
		\hline
		$ R_V^i $ & visual feature vector of post $p_i$ \\
		\hline
		$ R_G^i $ & social context feature vector of post $p_i$ \\
		\hline
		$ R^i $ & initial fused feature vector of post $p_i$ \\
		\hline
		$ Z_T^i $ & textual feature vector obtained by applying multi-head self-attention on $ R_T^i $ \\
		\hline
		$ Z_V^i $ & visual feature vector obtained by applying multi-head self-attention on $ R_V^i $ \\
		\hline
		$ Z_{TV}^i $ & output feature vector obtained by applying co-attention between $ Z_T^i $ and $ Z_V^i $ \\
		\hline
		$ Z_{VT}^i $ & output feature vector obtained by applying co-attention between $ Z_V^i $ and $ Z_T^i $ \\
		\hline
		$ Z^i $ & final feature vector of the intrinsic modality for post $ p^i $ \\
		\hline
		$ X_{fuse}^i $ & final feature vector of the fused intrinsic and social modalities for post $ p^i $ \\
		\hline
		$ \mathcal{L}_{scl} $ & supervised contrastive loss \\
		\hline
		$ \mathcal{L}_{cmca} $ & cross-modal consistency alignment loss \\
		\hline
		$ \mathcal{L}_{ml} $ & mutual learning loss \\
		\hline
		$ \mathcal{L}_{ce} $ & cross-entropy loss \\
		\hline
		$ \mathcal{L}_{overall} $ & overall loss \\
		\hline
	\end{tabular}
	\label{tab:notations}
\end{table}

\section{The Proposed Method}\label{sec:mtd}
\subsection{Framework}
\begin{figure*}[t]  
	\centering  
	\captionsetup{font=small}
	\includegraphics[width=0.9\textwidth]{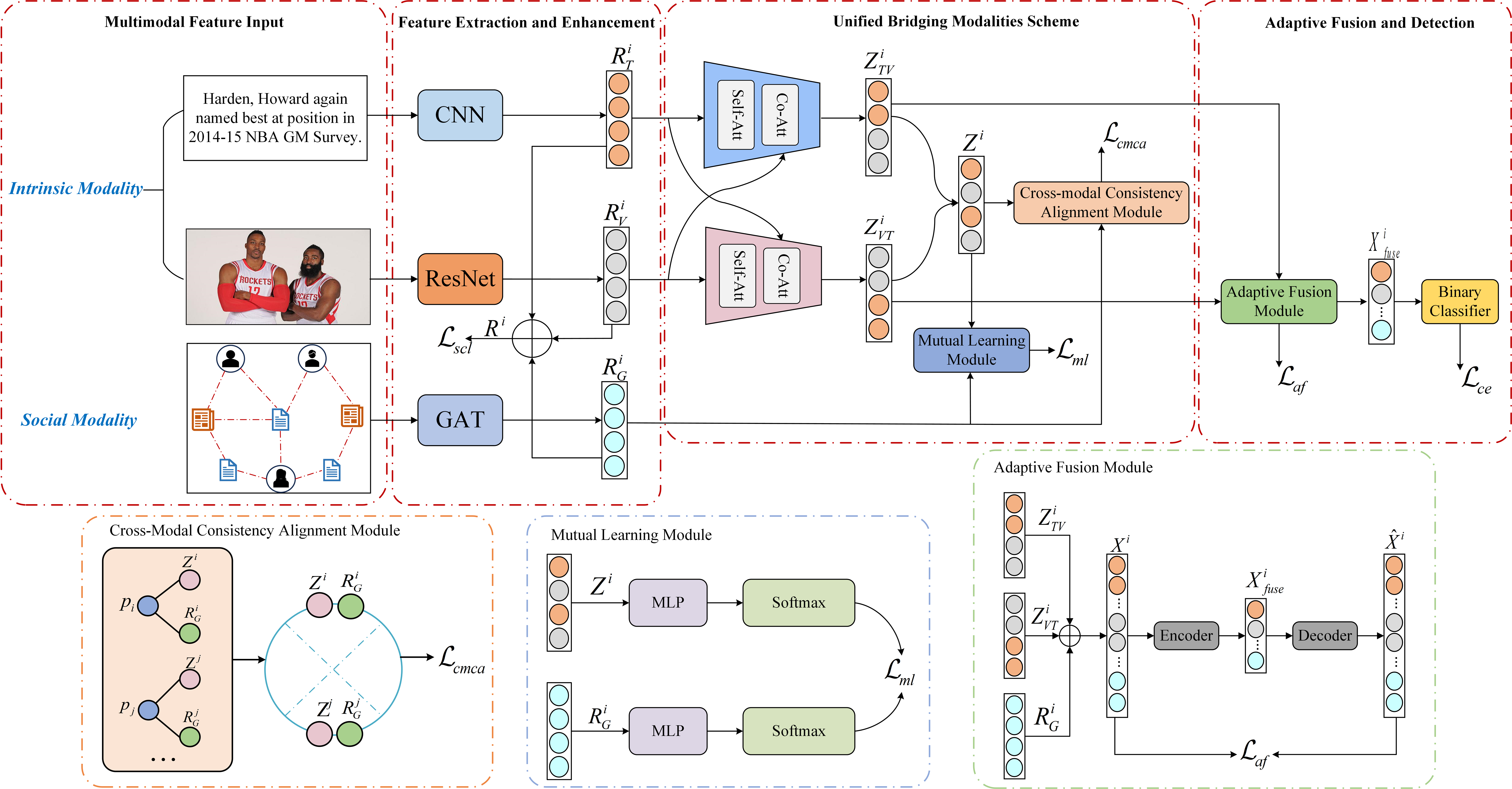} 
	\caption{The proposed ISMAF framework. The above is an overview structure and the internal structures of three modules are below. The intrinsic and social modality information of a post is initially fed into their respective feature extractors. Subsequently, the initial representations are refined through supervised contrastive learning. ISMAF then captures the correspondence between intrinsic and social modalities through the \textit{unified bridging modalities scheme}, which incorporates the \textit{cross-modal consistency alignment module} and the \textit{mutual learning module}. Finally, the intrinsic and social modalities are fused by an \textit{adaptive fusion module}, followed by a binary classifier to generate a prediction.}  
	\label{fig1}
	\vspace{-10pt}
\end{figure*}

The proposed multimodal rumor detection framework, namely Intrinsic-Social Modality Alignment and Fusion (ISMAF), is shown in Fig.~\ref{fig1}. The framework consists of three key components: \textit{feature extraction and enhancement}, \textit{unified bridging modalities scheme}, and \textit{Adptive Fusion and Detection}, which together achieve effective rumor detection.

Given a post \(p_i\), the \textit{feature extraction and enhancement} component first extracts responsible for obtaining initial representations from both intrinsic modality (textual and visual feature) and social modality (social context feature) of \(p_i\). These initial representations are then refined using supervised contrastive learning, which enhances the quality of the embeddings. Next, the \textit{Unified Bridging Modalities Scheme} integrates two important modules: the \textit{Cross-Modal Consistency Alignment Module} and the \textit{Mutual Learning Module}. These modules work together to align the intrinsic modality closely with the corresponding social modality, effectively capturing the correspondence between them. Finally, the \textit{Adptive Fusion and Detection} component employs an adaptive fusion mechanism to effectively fuse textual, visual, and social context features into a unified representation \(X_{fuse}^i\). This fused representation is then used for the final classification prediction, ensuring that all relevant multimodal information is fully leveraged.

\subsection{Feature Extraction and Enhancement}
\subsubsection{Unimodal Feature Extraction} First, we perform feature extraction on each modality as follows:
\begin{itemize}
	\item \textbf{Textual Features:} Given a  text \( t_i \), its word embedding is represented as \( O^i \in \mathbb{R}^{L_t \times d} \) , following the same method in \cite{yuan2019jointly}, where \(L_t\) denotes the length of the text sequence, and
	\(d\) represents the embedding dimension of the text. Then the word embedding matrix \(O^i \) is processed through a CNN\cite{yu2017convolutional}, followed by max pooling to generate the feature map \(\hat{s_k^i}\), where \( k \) indicates the kernel size. We concatenate the pooled outputs from various receptive fields to obtain the final textual representation \(R_T^i  \in \mathbb{R}^d\).
	
	\item \textbf{Visual Features:} For each image \(v_i\), we utilize ResNet-50\cite{he2016deep} to extract features. The output from its second-to-last layer \(V_r^i\) is passed through a fully connected layer, producing the visual representation \(R_V^i\), which is of the same dimension as the textual representation.
	
	\item \textbf{Social Context Features:} To capture social contexts, we first construct a social graph where the nodes represent posts, comments, and users. The initial embeddings for posts and comments are derived from their text representations, while user embeddings are calculated as the average of their associated post and comment embeddings. The edges between nodes are formed based on cosine similarity, where a threshold determines the connections.
\end{itemize}
	
	Following \cite{zheng2022mfan}, we update the node embeddings using Graph Attention Networks (GAT) with the signed attention mechanism. A multi-head attention mechanism is then employed to enrich the features from multiple perspectives. The final social context representation \(R_G^i\) for the post \(p_i\) is derived from the updated graph features. To ensure consistency across all modalities, \(R_G^i\) is in the same dimensional space \(d\).
	
\subsubsection{Multimodal Representation Enhancement} 
Our goal in multimodal post representation is to create a feature space that enables accurate classification of posts as real or fake. To this end, we employ contrastive learning to enhance the model’s ability to distinguish between these classes across various modalities.

Specifically, our approach employs supervised contrastive learning (SCL)\cite{khosla2020supervised}, which enables the model to better align with the specific requirements of downstream tasks. Specifically, we concatenate three types of features to form the initial fused representation for the \(i\)-th sample, denoted as \(R^i=R_T^i\bigoplus R_V^i\bigoplus R_G^i \), where \(R^i \in \mathbb{R}^{3d} \), \(\bigoplus\) represents the concatenation operation. The supervised contrastive loss function is defined as:
\begin{equation}
	\mathcal{L}_{scl}=\sum_{i=1}^N \frac{1}{\left|P\left(i\right)\right|} \sum_{p \in P\left(i\right)}\log \frac{\exp \left(R^{i} \cdot R^{p} / \tau \right)}{\sum_{a \in A\left(i\right)} \exp \left(R^{i} \cdot R^{a} / \tau \right)}, \label{eq1}
\end{equation}
where \(N\) denotes the mini-batch size, \(P\left(i\right)\) contains the indices of positive samples within the batch, and \(\left|P\left(i\right)\right|\) represents the number of positive samples. The numerator captures the similarity between representations of samples sharing the same label, while the denominator accounts for the similarity across all samples. The temperature parameter \(\tau\) regulates the smoothness of the distribution.

The supervised contrastive loss function plays a crucial role in enhancing the multimodal representation by explicitly guiding the model to more effectively cluster representations of samples with the same label, while simultaneously pushing apart those of opposing labels. This approach reinforces the distinction between real and fake posts.

\subsection{A Unified Scheme for Bridging Modalities}
In the representation of three feature types, we categorize text and image features as intrinsic modality, while social context features are classified as the social modality. Traditional approaches often treat these feature types homogenously, overlooking the distinct and complementary nature of intrinsic and social modalities. This oversight hampers the exploration of intrinsic-social correspondence, which is essential for a comprehensive understanding of post content. To address this limitation, we propose a unified bridging scheme that harmoniously integrates these modalities, effectively capturing their correspondence and interplay. Specifically, this scheme incorporates a cross-modal consistency alignment module and a mutual learning method.

\subsubsection{Cross-Modal Consistency Alignment}
The cross-modal consistency alignment module begins by leveraging a multi-head attention mechanism to integrate information from post text and images, generating a comprehensive representation of the intrinsic modality. This facilitates interaction between textual and visual features, allowing the model to effectively capture complementary insights from the intrinsic modality.

Given a post \(p_i\) and its textual and visual features \(R_T^i\) and \(R_V^i\), we first apply multi-head self-attention to construct augmented representations \(Z_T^i\) and \(\hat{Z_V^i}\) as follows:

\begin{equation}
	Z_{m}^{i} = \bigoplus_{h=1}^{H} \left( \text{softmax}\left(\frac{Q_{m}^{i} (K_{m}^{i})^{T}}{\sqrt{d}}\right) V_{T}^{i} \right) W_{m}^{O},\label{eq2}
\end{equation}
where \(Z_{m}^{i}\) represents the output feature, and \(Q_{m}^{i}\), \(K_{m}^{i}\), \(V_{m}^{i}\) are the query, key, and value matrices, respectively, for the \(i\)-th instance in feature type \(m\), where \(m \in \{T,V\}\). The matrix \(W_{m}^{O} \in \mathbb{R}^{d*d}\) is the output of linear transformation, and \(\bigoplus_{h=1}^{H}\) represents the concatenation of all \(H\) attention heads.

We perform a co-attention process to fully leverage the interaction between the textual and visual features and enhance the representation of the intrinsic modality. This process generates two co-attention vectors, \(Z_{TV}^{i}\) and \(Z_{VT}^{i}\), which are then averaged to obtain the final intrinsic modality representation \(Z^i\) for the post \(p_i\), as shown below:

\begin{equation}
	Z_{TV}^{i} = \bigoplus_{h=1}^{H} \left( \text{softmax}\left(\frac{Q_{T}^{i} (K_{V}^{i})^{T}}{\sqrt{d}}\right) V_{V}^{i} \right) W_{TV}^{O}, \label{eq3}
\end{equation}

\begin{equation}
	Z_{VT}^{i} = \bigoplus_{h=1}^{H} \left( \text{softmax}\left(\frac{Q_{V}^{i} (K_{T}^{i})^{T}}{\sqrt{d}}\right) V_{T}^{i} \right) W_{VT}^{O}, \label{eq4}
\end{equation}

\begin{equation}
	Z^i = \frac{1}{2}\left(Z_{TV}^{i}+Z_{VT}^{i}\right). \label{eq5}
\end{equation}

This module is proposed to enhance the correlation between intrinsic and social modalities. To achieve this, we employ a contrastive learning approach proposed in \cite{WOS:000683178501065} that maximizes the similarity between \(Z^{i}\) and \(R_{G}^{i}\), which correspond to the same post, while minimizing the similarity with all other vectors within the mini-batch. This strategy encourages the intrinsic modality representations to be closely embedded with their corresponding social modality representations in the latent space. Such alignment reinforces cross-modal consistency and effectively bridges the gap between intrinsic and social modalities. The loss function \(\mathcal{L}_{cmce}\) for the positive pairs of intrinsic modality representations \(Z\) and social modality representations \(R_{G}\) can be calculated as:

\begin{equation}
	\scalebox{0.985}{$
		l\left(z, r\right) = -\log \frac{\exp\left(\frac{sim\left(z^i, r^i\right)}{\tau}\right)}
		{\sum\limits_{\substack{k=1 \\ k \neq i}}^{N} \exp\left(\frac{sim\left(z^i, z^k\right)}{\tau}\right) + \sum\limits_{k=1}^{N} \exp\left(\frac{sim\left(z^i, r^k\right)}{\tau}\right)}$}, \label{eq6}
\end{equation}

\begin{equation}
	\mathcal{L}_{cmca} = \frac{1}{2N}\sum_{i=1}^{N}\left(l\left(Z, R_G\right)+l\left(R_G, Z\right)\right), \label{eq7}
\end{equation}
where \(N\) is the mini-batch size, \(sim\left(\cdot\right)\) represents the cosine similarity function and \(\tau\) is the temperature parameter.

\subsubsection{Mutual Learning}
We utilize a collaborative learning method to facilitate mutual learning between classifiers based on intrinsic and social modalities. This method leverages a logistic distribution loss function to measure the Kullback-Leibler (KL) discrepancy between two classifiers, thereby enabling effective knowledge transfer and improving the accuracy of detection.

Initially, intrinsic modality representations \(Z\) and social modality representations \(R_G\) are individually projected into a common latent space through the transformation as follows:

\begin{equation}
	E_Z = ReLU(ZW_Z+b_Z), \label{eq8}
\end{equation}

\begin{equation}
	E_{R_G} = ReLU(R_{G}W_{R_G}+b_{R_G}), \label{eq9}
\end{equation}
where \(E_Z\) and \(E_{R_G}\) represent the mapped intrinsic and social semantics in the common latent space, respectively.

To derive the probability distribution of the category labels \(P_Z\) and \(P_{R_G}\), the transformed semantics \(E_Z\) and \(E_{R_G}\) are processed through a fully connected (FC) layer, followed by a softmax function:

\begin{equation}
	P_Z = softmax(FC(E_Z)), \label{eq10}
\end{equation}

\begin{equation}
	P_{R_G} = softmax(FC(E_{R_G})). \label{eq11}
\end{equation}

The mutual learning process is driven by the optimization of the KL divergence between the probability distributions \(P_Z\) and \(P_{R_G}\). This process fosters collaboration between the classifiers, enabling them to complement each other by capturing correlated information across the intrinsic and social modalities. By minimizing the KL divergence between their respective probability distributions, the learning dynamics of the two classifiers are aligned, which enhances their generalization capabilities. Ultimately, this method leads to a more robust understanding of the underlying data and contributes to overall performance. The KL divergence is defined as:

\begin{equation}
	KL\left(P||Q\right) = \sum_{i=1}^{N} P\left(i\right) \log\frac{P\left(i\right)}{Q\left(i\right)}. \label{eq12}
\end{equation}

Consequently, the mutual learning loss function \(\mathcal{L}_{ml}\) can be formulated as:
\begin{equation}
	\mathcal{L}_{ml} = \frac{1}{2}\left(KL\left(P||Q\right)+KL\left(Q||P\right)\right). \label{eq13}
\end{equation}

\subsection{Adaptive Fusion Mechanism}
Due to the complex and diverse relationships among the three modalities, traditional fusion methods often fall short of capturing their intricate interdependencies. Simple concatenation methods treat each modality as independent, overlooking the subtle interactions between them, which results in suboptimal feature integration. Similarly, attention-based approaches, while capable of inferring inter-modal interactions, are restricted by fixed attention patterns that cannot dynamically adjust to the varying relevance of each modality. As a result, these methods struggle to fully exploit the complementary information intrinsic in each modality, leading to poor representations and mediocre performance in downstream tasks.

To address these limitations, we introduce an adaptive fusion mechanism inspired by auto-fusion techniques \cite{sahu2019adaptive}. This mechanism dynamically adjusts to the varying contributions of each modality, ensuring that the relevant information from both intrinsic and social modalities is effectively integrated. By employing an encoder-decoder structure, we achieve effective compression of the multimodal data into a unified representation while capturing complex interrelationships among features. The encoder transforms the concatenated intrinsic and social modality representation into a unified representation., learning a lower-dimensional representation that retains essential information. Subsequently, the decoder reconstructs this representation, preserving the integrity of the original multimodal information.

Specifically, for a given post \(p_i\), we first concatenate intrinsic modality representation \(Z_{TV}^i\), \(Z_{VT}^i\), and social modality representation \(R_{G}^i\), obtaining a composite vector \(X^i=Z_{TV}^i\bigoplus Z_{VT}^i\bigoplus R_{G}^i\), where \(X^i \in \mathbb{R}^{3d}\) . Then, we compress the vector \(X^i\) using an encoder to reduce its dimensionality to \(d\), yielding a vector \(X_{fuse}^i\). Next, we employ a decoder to reconstruct \(X_{fuse}^i\) producing the recovered vector \(\hat{X^i}\). The reconstruction loss is defined as:

\begin{equation}
	\mathcal{L}_{af} = \left\|\hat{X}^{i}-X^{i}\right\|^{2}. \label{eq14}
\end{equation}

We treat the intermediate vector \(X_{fuse}^i\) as the fused multimodal representation which is then used for rumor detection.

\subsection{Classification}
The final multimodal representation \(X_{fuse}^i\) of each post \(p_i\) is passed through a fully connected layer followed by a softmax function to predict its classification as either a rumor or not. The prediction is given by:

\begin{equation}
	\hat{y_i} = softmax\left(FC\left(X_{fuse}^i\right)\right), \label{eq15}
\end{equation}
where \(\hat{y_i}\) represents the predicted probability that post \(p_i\) s classified as a rumor. We use the cross-entropy loss function for classification, that is:
\begin{equation}
	\mathcal{L}_{ce} = \sum_{i=1}^{N}y_i\log\left(\hat{y_i}\right)+\left(1-y_i\right)\log\left(1-\hat{y_i}\right). \label{eq16}
\end{equation}

The overall loss function combines multiple components is:
\begin{equation}
	\mathcal{L}_{overall} = \mathcal{L}_{ce} + \lambda_1 \mathcal{L}_{scl} + \lambda_2 \mathcal{L}_{cmca} + \lambda_3 \mathcal{L}_{ml} + \lambda_4 \mathcal{L}_{af}, \label{eq17}
\end{equation}
where hyperparameters \(\lambda_1, \lambda_2, \lambda_3, \lambda_4\) are used to weight the importance of each loss component.

\begin{algorithm}
	\small
	\caption{ISMAF Algorithm}\label{alg:algorithm1}
	\begin{algorithmic}[1]
		\REQUIRE Training set $P = \{p_1,p_2,\cdots,p_N\}$, where each $p_i$ consists of text, visual, and social graph data; mini-batch size $B$; number of epochs $\epsilon$
		\ENSURE A binary classifier model, $f(p_i) \rightarrow y$
		\STATE Initialize epoch counter $e = 0$
		\FOR{$e = 1$ to $\epsilon$}
		\FOR{each mini-batch in $\lceil \frac{N}{B} \rceil$ iterations}
		\STATE Randomly sample a mini-batch from $P$
		\STATE Extract textual feature $R_T^i$ using CNN
		\STATE Extract visual feature $R_V^i$ using ResNet50
		\STATE Extract social context feature $R_G^i$ using GAT
		\STATE Concatenate features: $R^i = R_T^i \bigoplus R_V^i \bigoplus R_G^i$
		\STATE Compute $\mathcal{L}_{scl}$ using $R^i$ by \eqref{eq1}
		\STATE Obtain augmented features $Z_{TV}^i$ and $Z_{VT}^i$ by \eqref{eq2}--\eqref{eq4}
		\STATE Derive intrinsic feature: $Z^i = \frac{1}{2}(Z_{TV}^i + Z_{VT}^i)$
		\STATE Compute $\mathcal{L}_{cmca}$ using $Z^i$ by \eqref{eq6}--\eqref{eq7}
		\STATE Calculate label distributions $P_Z$ and $P_{R_G}$ by \eqref{eq8}--\eqref{eq11}
		\STATE Compute $\mathcal{L}_{ml}$ by \eqref{eq12} and \eqref{eq13}
		\STATE Concatenate features: $X^i = Z_{TV}^i \bigoplus Z_{VT}^i \bigoplus R_{G}^i$
		\STATE Obtain fused vector $X_{fuse}^i \leftarrow \text{Encoder}(X^i)$; recover vector $\hat{X}^i \leftarrow \text{Decoder}(X_{fuse}^i)$
		\STATE Compute $L_{af}$ using $X^i$ and $\hat{X}^i$ by \eqref{eq14}
		\STATE Predict label $\hat{y}_i$ and compute $L_{ce}$ by \eqref{eq15}--\eqref{eq16}
		\STATE Calculate overall loss: $\mathcal{L}_{overall} = \mathcal{L}_{ce} + \lambda_1 \mathcal{L}_{scl} + \lambda_2 \mathcal{L}_{cmca} + \lambda_3 \mathcal{L}_{ml} + \lambda_4 \mathcal{L}_{af}$
		\STATE Backpropagate and update model parameters
		\ENDFOR
		\ENDFOR
		\RETURN The trained model, $f(p_i) \rightarrow y$
	\end{algorithmic}
\end{algorithm}

\subsection{Algorithm}
We list the whole procedure of the proposed ISMAF method in Algorithm~\ref{alg:algorithm1}. Given a training set $P = \{p_1,p_2,\cdots,p_N\}$ where each $p_i$ consists of text, visual, and social graph data, our objective is to train a binary classifier that learns a function \( f(p_i) \rightarrow y \), mapping each post \( p_i \) to its corresponding label.

In the initial steps of the algorithm, parameters such as the epoch counter and mini-batch size are initialized. Lines 5 to 7 focus on extracting the initial textual, visual, and social context features. Subsequently, in lines 8 and 9, the supervised contrastive loss $\mathcal{L}_{scl}$ is computed using the concatenated feature representations $R^i$. Lines 10 to 12 detail the computation of the cross-modal consistency loss $\mathcal{L}_{cmca}$, which is based on the intrinsic feature vector $Z^i$. The mutual learning loss $\mathcal{L}_{ml}$ is then derived in lines 13 and 14, leveraging label distributions generated from both intrinsic modality and social graph features. In lines 15 and 16, the fused vector $X_{fuse}^i$ is obtained through an encoder-decoder structure, and the adaptive fusion loss $\mathcal{L}_{af}$ is calculated. Finally, the overall loss function $\mathcal{L}_{overall}$ is formulated as a weighted combination of these losses, followed by backpropagation to iteratively update the model parameters.

\section{Experiments}\label{sec:exp}

\subsection{Experimental Settings}
\paragraph{Datasets}  Given the nature of our research, which involves leveraging text, image, user, and comment information, we carefully selected two real-world datasets as representative benchmarks: Weibo\cite{song2019ced} for Chinese social media and PHEME\cite{yu2017convolutional} for English social media, consistent with the datasets used in \cite{zheng2022mfan} and \cite{xu2024clffrd}.

The Weibo dataset, collected from Sina Weibo, one of China’s most popular social media platforms, and the PHEME dataset, which consists of tweets from five major breaking news events on Twitter, both provide rich multimodal data, encompassing textual, visual, and social context information. To ensure data quality and alignment with our research objectives, we annotated both datasets with binary labels (Rumors and Non-Rumors) and excluded instances missing either text or image data during preprocessing. The statistics of these cleaned datasets are summarized in Table \ref{tab1}.

\begin{table}[htbp]
	\begin{center}
		\caption{The Dataset statistics}
		\begin{tabular}{ccccc}
			\toprule
			\textbf{Datasets} & \textbf{Figures} & \textbf{Users} & \textbf{Comments} & \textbf{Labels} \\ 
			\midrule
			\textbf{PHEME}   & 2018   & 894  & 7388  & N: 1428 / R: 590  \\
			\textbf{Weibo}   & 1467   & 985  & 4534  & N: 877 / R: 590   \\
			\bottomrule
			\multicolumn{5}{r}{N: Non-rumors; R: Rumors.}
		\end{tabular}
		\label{tab1}
	\end{center}
\end{table}

\paragraph{Parameter Settings}
Each dataset is divided into training, validation, and test sets with a 70\%-10\%-20\% split. The number of attention heads $H$ is set to 8, and the feature dimension $d$ is set to 300. Following \cite{zheng2022mfan}, Projected Gradient Descent (PGD)\cite{2017Towards} is applied during model training. The Adam optimizer\cite{2014Adam} is employed with a dynamic learning rate decay, initialized at 0.002. Additionally, dropout with a rate of 0.5 is applied to mitigate overfitting. The method is implemented in PyTorch and evaluated on a Linux server equipped with an RTX 4080 GPU. All experiments are conducted five times, and the reported results represent the average performance across these runs.

\paragraph{Evaluation Metric} We utilize four widely used evaluation metrics: Accuracy  (ACC), Precision (Pre), Recall (Rec), and F1-score (F1) to evaluate the performance of our proposed framework as well as other baseline models. These metrics provide a comprehensive assessment of the methods.

\subsection{Methods in Comparison} 
We compare it with several state-of-the-art methods, which are briefly summarized as follows:

\begin{itemize}
	\item \textbf{QSAN} \cite{tian2020qsan} combines quantum-driven text encoding with a signed attention mechanism, allowing for a more nuanced analysis of textual data in the context of fake information detection.
	
	\item \textbf{EANN} \cite{wang2018eann} employs a GAN-based approach to integrate text and image data, focusing on extracting invariant features across different events.
	
	\item \textbf{MVAE} \cite{khattar2019mvae} utilizes a multimodal variational autoencoder to learn shared representations of text and image data, enabling the model to exploit the complementary information present in both modalities.
	
	\item \textbf{SAFE} \cite{zhou2020safe} leverages multimodal features and cross-modal similarity to effectively analyze and debunk fake news. The model effectively aligns textual and visual information to enhance its accuracy.
	
	\item \textbf{EBGCN} \cite{wei2021towards} applies a Bayesian approach to model the reliability of relationships within propagation structures, capturing the dynamics of information diffusion. 
	
	\item \textbf{GLAN} \cite{yuan2019jointly} employs an attention-based network to integrate local semantic and global structural information derived from a heterogeneous graph representation of social context, thereby enhancing rumor detection performance.
	
	\item \textbf{MFAN} \cite{zheng2022mfan} integrates textual, visual, and social context features within a feature-enhanced attention network for rumor detection. By exploiting the hidden connections in social network data, MFAN effectively leverages social context to improve detection performance.
	
	\item \textbf{CLFFRD} \cite{xu2024clffrd} introduces a novel approach to rumor detection, incorporating curriculum learning to address sample difficulty and a fine-grained fusion strategy to enhance feature integration. 
\end{itemize}

The methods used in comparison can be categorized based on their feature utilization.  QSAN relies solely on textual features. EANN, MVAE, and SAFE incorporate both textual and visual features. EBGCN and GLAN leverage social context features, emphasizing structural information in their analysis. MFAN and CLFFRD employ a comprehensive approach by integrating textual, visual, and social context features, and have demonstrated superior performance, representing the state-of-the-art methods in the field.

\subsection{Experimental Results and Analysis}
Table \ref{tab2} presents the average performance and standard deviation across five runs on the PHEME and Weibo datasets. As illustrated, our ISMAF model consistently outperforms the baseline methods across all evaluation metrics on both datasets. Key observations are as follows:

In comparison to multimodal approaches, QSAN, a text-only method, struggles to capture the visual and contextual cues inherent in multimodal data. Consequently, its performance on a dataset rich in multimodal content is significantly inferior to that of multimodal models, highlighting the limitations of relying solely on text-based features.

Methods such as EANN, MVAE, and SAFE enhance rumor detection by incorporating both textual and visual information. Among them, SAFE is particularly effective in fusing these modalities. Nevertheless, these models fail to fully leverage the contextual information embedded in social interactions and propagation structures of posts. This oversight limits their ability to capture the broader context surrounding rumors and fake news, potentially hindering their performance in more complex scenarios.

In contrast, EBGCN and GLAN emphasize the importance of social context features, which are crucial for understanding the propagation and structural dynamics of misinformation. However, their limited integration of textual and visual information can constrain their overall performance. In comparison, ISMAF provides a more comprehensive framework by integrating social context with both textual and visual modalities, which allows ISMAF to capture both the structural dynamics of social interactions and the content-based features from text and images, leading to superior performance.

Multimodal approaches such as MFAN and CLFFRD leverage textual, visual, and social context features to improve performance by utilizing richer information sources. However, ISMAF further advances the state-of-the-art by addressing two key challenges. First, it explores the often-neglected interplay between intrinsic and social modalities, capturing complementary insights that deepen the understanding of news content. Second, ISMAF overcomes the inherent difficulties in fusing three modalities, a challenge that many existing methods struggle to address. Through its innovative adaptive fusion mechanism, ISMAF effectively integrates text, image, and social context, leading to superior performance.

\begin{table*}[htbp]
	\caption{Comparison with the State-of-the-Art Methods on Two Datasets}
	\centering
	\resizebox{\textwidth}{!}{
		\begin{tabular}{c c c c c c c c c}
			\toprule
			\multirow{2}{*}{\textbf{Method}} & \multicolumn{4}{c}{\textbf{PHEME}} & \multicolumn{4}{c}{\textbf{Weibo}} \\
			\cmidrule(lr){2-5} \cmidrule(lr){6-9}
			& \textbf{ACC} & \textbf{Pre} & \textbf{Rec} & \textbf{F1} & \textbf{ACC} & \textbf{Pre} & \textbf{Rec} & \textbf{F1} \\
			\midrule
			QSAN \cite{tian2020qsan} &  $75.13 \pm 1.19$  &  $69.97 \pm 2.03$  &  $65.80 \pm 1.72$  &  $66.87 \pm 1.70$  &  $71.01 \pm 1.81$  &  $71.02 \pm 0.95$  &  $67.54 \pm 3.27$  &  $67.58 \pm 3.59$  \\
			EANN \cite{wang2018eann} & $77.13 \pm 0.96$  &  $71.39 \pm 1.07$  &  $70.07 \pm 2.19$  &  $70.44 \pm 1.69$  &  $80.96 \pm 2.26$  &  $80.19 \pm 2.37$  &  $79.68 \pm 2.46$  &  $79.87 \pm 2.40$  \\
			MVAE \cite{khattar2019mvae} & $77.62 \pm 0.64$  &  $73.49 \pm 0.81$  &  $72.25 \pm 0.90$  &  $72.77 \pm 0.81$  &  $71.67 \pm 0.89$  &  $70.52 \pm 0.95$  &  $70.21 \pm 1.01$  &  $70.34 \pm 0.98$  \\
			SAFE \cite{zhou2020safe} & $81.49 \pm 0.84$  &  $79.88 \pm 1.22$  &  $79.50 \pm 0.81$  &  $79.68 \pm 0.70$  &  $84.95 \pm 0.85$  &  $84.98 \pm 0.82$  &  $84.95 \pm 0.91$  &  $84.96 \pm 0.86$  \\
			EBGCN \cite{wei2021towards} & $82.99 \pm 0.65$  &  $81.31 \pm 0.73$  &  $79.29 \pm 0.71$  &  $79.82 \pm 0.64$  &  $83.14 \pm 2.01$  &  $85.46 \pm 2.12$  &  $81.76 \pm 1.54$  &  $81.45 \pm 1.74$  \\
			GLAN \cite{yuan2019jointly} & $83.32 \pm 1.64$  &  $81.25 \pm 2.06$  &  $77.13 \pm 3.26$  &  $78.51 \pm 2.68$  &  $82.44 \pm 2.02$  &  $82.45 \pm 2.26$  &  $80.86 \pm 1.71$  &  $81.26 \pm 1.93$  \\
			MFAN \cite{zheng2022mfan} & $88.73 \pm 0.83$  &  $87.07 \pm 1.41$  &  $85.61 \pm 1.65$  &  $86.16 \pm 1.04$  &  $88.95 \pm 1.43$  &  $88.91 \pm 1.60$  &  $88.13 \pm 1.68$  &  $88.33 \pm 1.53$  \\
			CLFFRD \cite{xu2024clffrd} & $89.95 \pm 0.73$  &  $88.26 \pm 0.86$  &  $87.57 \pm 0.74$  &  $88.13 \pm 0.77$  &  $91.26 \pm 1.24$  &  $90.23 \pm 1.29$  &  $89.70 \pm 1.24$  &  $89.82 \pm 1.28$  \\
			\textbf{ISMAF (ours)} & \textbf{\bm{$91.01 \pm 0.57$}} & \textbf{\bm{$89.46 \pm 1.19$}} & \textbf{\bm{$88.88 \pm 1.26$}} & \textbf{\bm{$89.09 \pm 0.69$}} & \textbf{\bm{$93.42 \pm 0.70$}} & \textbf{\bm{$93.69 \pm 0.99$}} & \textbf{\bm{$92.39 \pm 0.62$}} & \textbf{\bm{$92.95 \pm 0.73$}} \\
			\bottomrule
	\end{tabular}}
	\label{tab2}
\end{table*}

\subsection{Ablation Experiments}
\subsubsection{Ablation Study of Important Components.} To evaluate the contributions of each component within the ISMAF model, we conduct an ablation study by systematically removing key modules as follows:

\begin{itemize}
	\item \textbf{\textit{w/o} MRE}: Excludes the multimodal representation enhancement module.
	\item \textbf{\textit{w/o} CMCA}: Removes the cross-modal consistency alignment module.
	\item \textbf{\textit{w/o} ML}: Omits the mutual learning process between intrinsic and social modalities.
	\item \textbf{\textit{w/o} CMCA\&ML}: Jointly excludes both the cross-modal consistency alignment strategy and mutual learning.
	\item \textbf{\textit{w/o} AF}: Replaces the adaptive fusion mechanism with a simple concatenation operation.
\end{itemize}

As shown in Table \ref{tab3}, all ablation variants exhibit a reduction in performance compared to the complete ISMAF model across both datasets, underscoring the importance of each component in its overall effectiveness. Specifically, excluding the multimodal representation enhancement (MRE) module results in a modest decline, with accuracy decreasing by 1.05\% on the PHEME dataset and by 1.67\% on the Weibo dataset. This suggests that while the MRE module contributes to the enhancement of multimodal features, its absence does not result in a drastic performance decrease. 

In contrast, the removal of either cross-modal consistency alignment (CMCA) or mutual learning (ML) leads to a more significant decline in performance. For example, on the PHEME dataset, excluding CMCA results in a 1.57\% drop in accuracy, while on Weibo, it drops by 1.22\%. These findings emphasize the essential roles of CMCA and ML in maintaining consistency and refining feature representations across modalities. The most substantial performance drop occurs when both CMCA and ML are removed. In this case, the accuracy on PHEME decreases by 2.06\%, and on Weibo by 2.57\%, underscoring the superior combined effect of these two components when used together. Their complementary roles enhance the integration and refinement of features across modalities, leading to a synergistic improvement in performance compared to when either component is used independently.

Finally, the inclusion of the adaptive fusion (AF) mechanism results in a notable performance boost across both datasets, further validating the effectiveness of our fusion strategy in improving detection accuracy.

\begin{table}[htbp]
	\caption{Ablation Study of Important Components}
	\centering
	\begin{tabular}{c c c c c}
		\toprule
		\textbf{Dataset} & \textbf{Method} & \textbf{ACC} & \textbf{F1} \\
		\midrule
		\multirow{6}{*}{\textbf{PHEME}} 
		& \raggedleft \textbf{ISMAF} & \bm{$91.01$} & \bm{$89.09$} \\
		& \raggedleft \textit{w/o} MRE & $89.96$ & $87.38$ \\
		& \raggedleft \textit{w/o} CMCA & $89.44$ & 87.23 \\
		& \raggedleft \textit{w/o} ML & $90.30$ & $88.14$ \\
		& \raggedleft \textit{w/o} CMCA\&ML & $88.95$ & $86.80$ \\
		& \raggedleft \textit{w/o} AF & $89.30$ & $86.96$ \\
		\midrule
		\multirow{6}{*}{\textbf{Weibo}} 
		& \raggedleft \textbf{ISMAF} & \bm{$93.42$} & \bm{$92.95$} \\
		& \raggedleft \textit{w/o} MRE & $91.75$ & $90.94$ \\
		& \raggedleft \textit{w/o} CMCA & $92.20$ & $91.55$ \\
		& \raggedleft \textit{w/o} ML & $91.19$ & $90.51$ \\
		& \raggedleft \textit{w/o} CMCA\&ML & $90.85$ & $90.16$ \\
		& \raggedleft \textit{w/o} AF & $91.86$ & $91.24$ \\ 
		\bottomrule
	\end{tabular}
	\label{tab3}
\end{table}

\subsubsection{Ablation Study with Alternative Fusion Strategies.} Despite their popularity in multimodal fusion, traditional concatenation and attention-based methods often fall short in capturing complex interactions between text, visual, and social context features. To assess the advantages of our proposed Adaptive Fusion (AF) mechanism in handling these multimodal interactions, we conduct a comparative ablation study, replacing AF with established fusion approaches. Specifically, we examine four alternative fusion strategies as follows:

\begin{itemize}
	\item \textbf{IS-concat}: This method concatenates the intrinsic representation \(Z^i\) and the social context representation \(R^i\), allowing for straightforward feature integration.
	\item \textbf{IS-att}: Here, an attention mechanism is applied to align and fuse the intrinsic and social modalities, aiming to enhance selective focus across features.
	\item \textbf{IS-co}: Building upon the co-attention mechanism in \cite{wu2021multimodal}, we employ two co-attention operations to capture interactions between the intrinsic representation \(Z^i\) and the social context representation \(R^i\), aiming to effectively integrate these modalities within our fusion framework.
	\item \textbf{Cross-co}: Based on \cite{zheng2022mfan}, this method incorporates a cross-modal co-attention mechanism, jointly processing the textual (\(R_T^i\)), visual (\(R_V^i\)), and social context (\(R_G^i\)) representations to address cross-modal dependencies.
\end{itemize}

As presented in Fig.~\ref{fig2}, we observe the following key insights from the experimental results:

\begin{figure}[t]
	\centering
	\captionsetup[subfloat]{font=small}
	\captionsetup{font=small}
	\subfloat[PHEME dataset]{
		\includegraphics[width=0.23\textwidth]{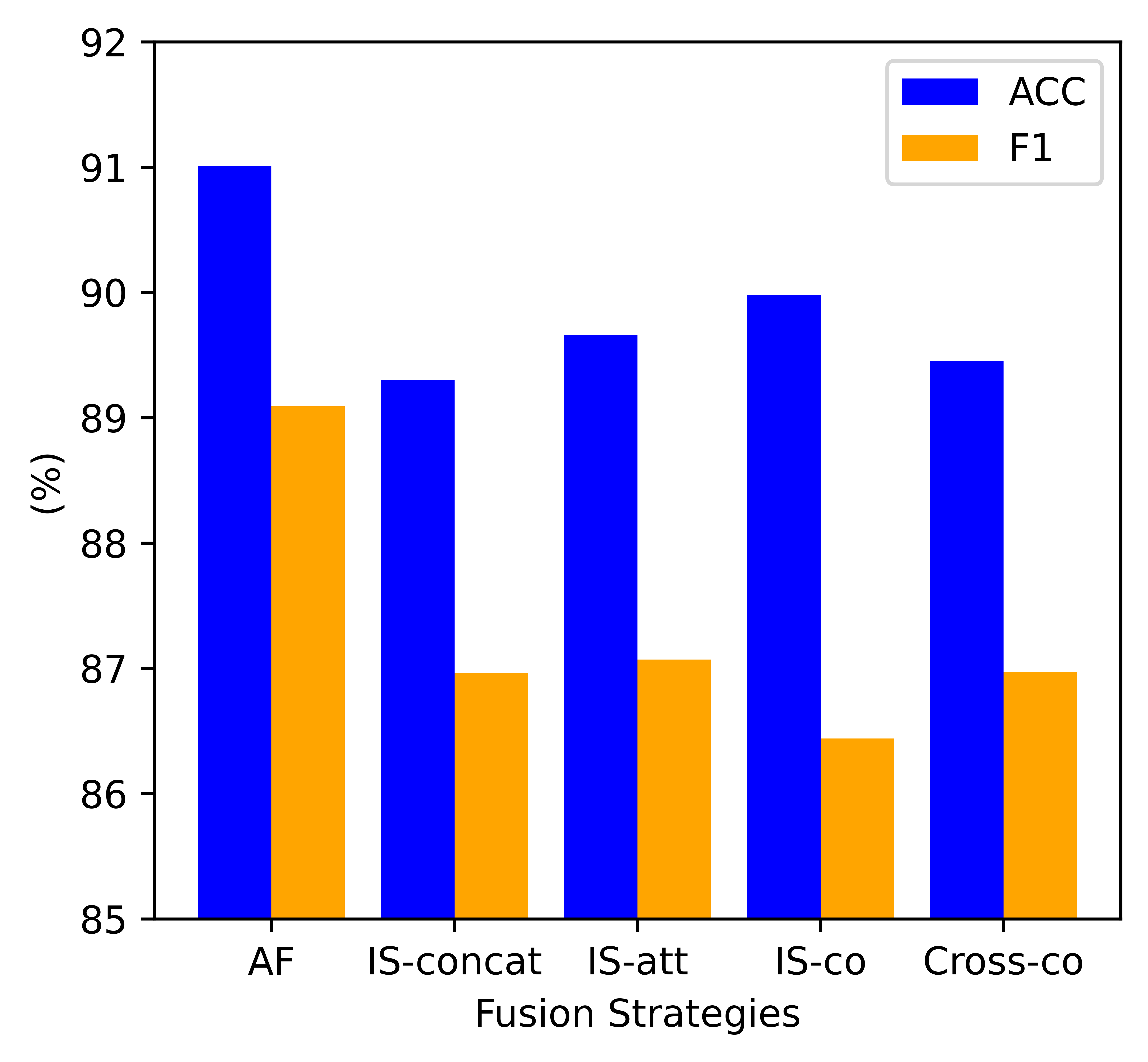}
		\label{subfig2_1}
	}
	\hfill
	\subfloat[Weibo dataset]{
		\includegraphics[width=0.23\textwidth]{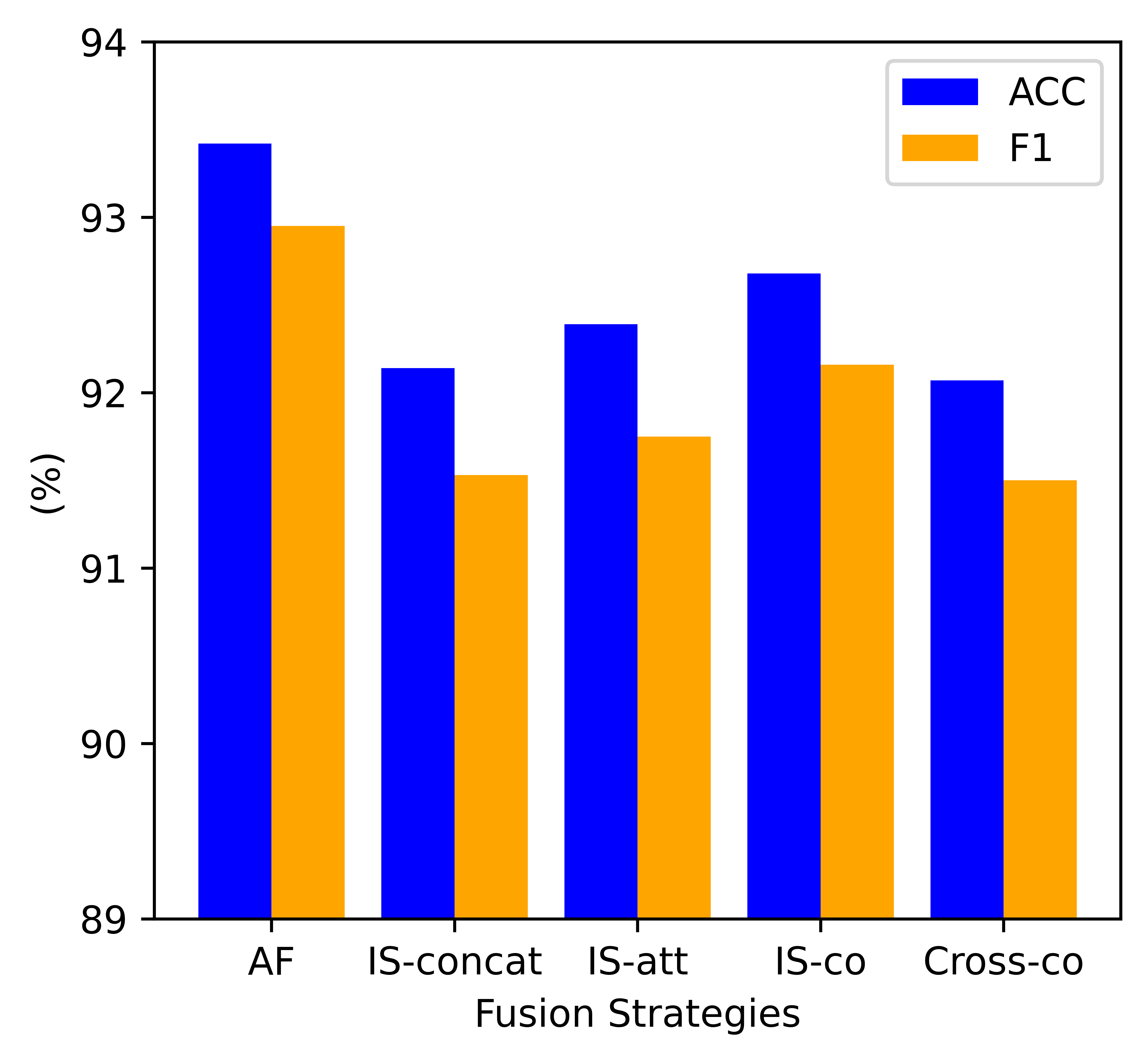}
		\label{subfig2_2}
	}
	\caption{Performance comparison of fusion strategies on the PHEME and Weibo datasets, illustrating the accuracy (ACC) and F1-score (F1) achieved by each method. This figure highlights the performance differences among various fusion strategies, including AF, IS-concat, IS-att, IS-co, and Cross-co.}
	\label{fig2}
	\vspace{-10pt}
\end{figure}

\begin{itemize}
	\item The Cross-co method yields suboptimal results across both datasets, suggesting that indiscriminately integrating textual, visual, and social context features may lead to information loss and fails to fully leverage the rich latent information inherent in each modality. In contrast, the other methods that process the intrinsic (textual and visual) and social (social context) modalities separately exhibit better performance, highlighting the advantage of distinguishing between these modality types.
	
	\item The IS-concat method, while straightforward, tends to overlook the complex interdependencies among modalities, which limits its ability to achieve an effective fusion of features. By simply concatenating intrinsic and social representations, this method lacks the mechanism to capture nuanced relationships, which are essential for richer and more informative feature integration.
	
	\item Both IS-att and IS-co provide improved integration over concatenation by enabling some level of modality interaction. The IS-att mechanism introduces interaction through fixed attention patterns, yet its rigidity may constrain the model's ability to effectively capture complementary information. The IS-co method, with its co-attention operations, partially addresses this limitation by capturing the complex interplay between intrinsic and social modalities; however, it still falls short of fully adapting to the unique characteristics of each modality. 
	
	\item In contrast, The AF method dynamically adapts fusion patterns to capture nuanced relationships among modalities. By selectively integrating relevant features through an encoder-decoder structure, AF achieves a more effective fusion of textual, visual, and social information, resulting in superior performance.
\end{itemize}

\subsection{Sensitivity Analysis}
\begin{figure*}[t]
	\centering
	\captionsetup[subfloat]{font=small}
	\captionsetup{font=small}
	\subfloat[]{
		\includegraphics[width=0.20\textwidth]{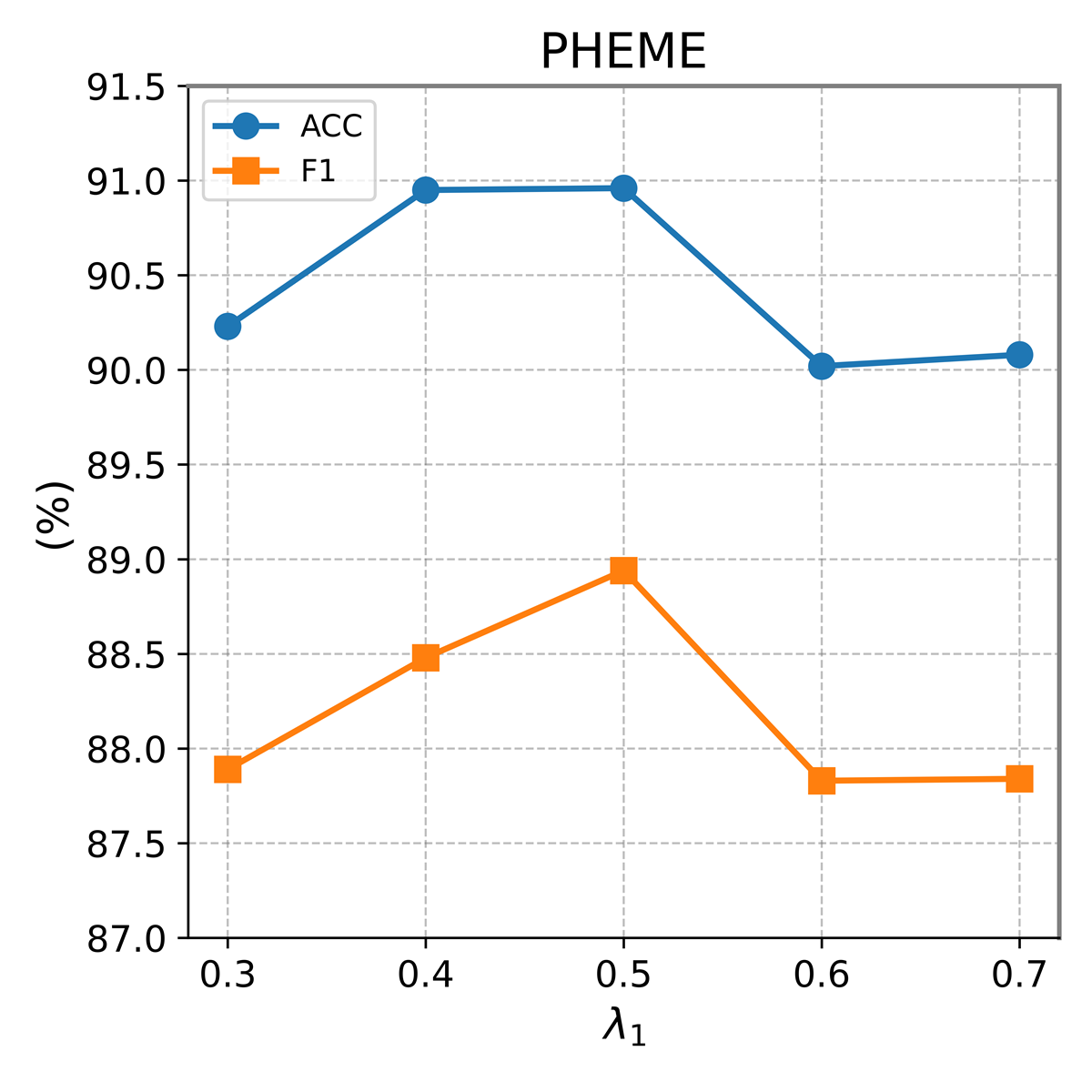}
		\label{subfig3_1}
	}
	\hfill
	\subfloat[]{
		\includegraphics[width=0.20\textwidth]{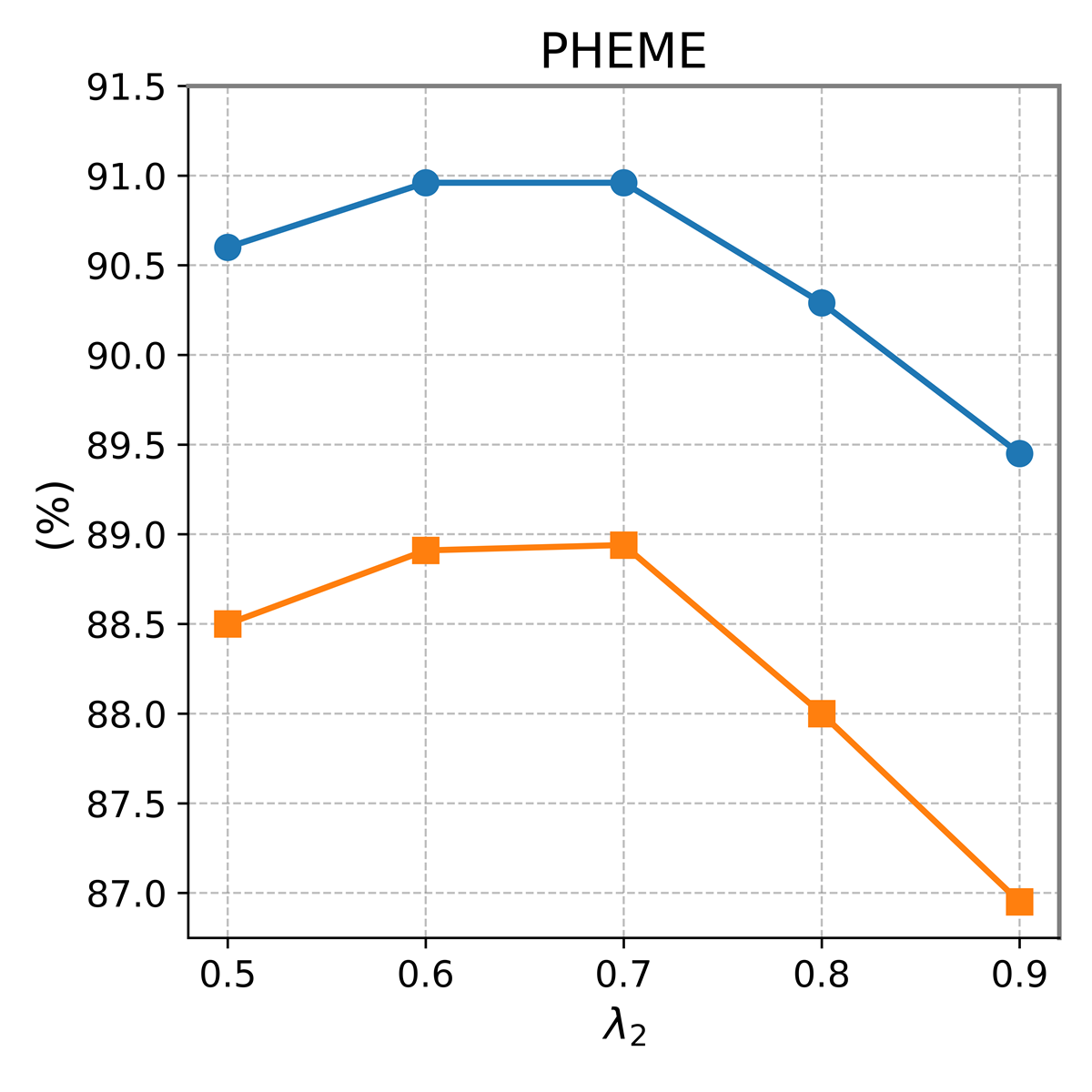}
		\label{subfig3_2}
	}
	\hfill
	\subfloat[]{
		\includegraphics[width=0.20\textwidth]{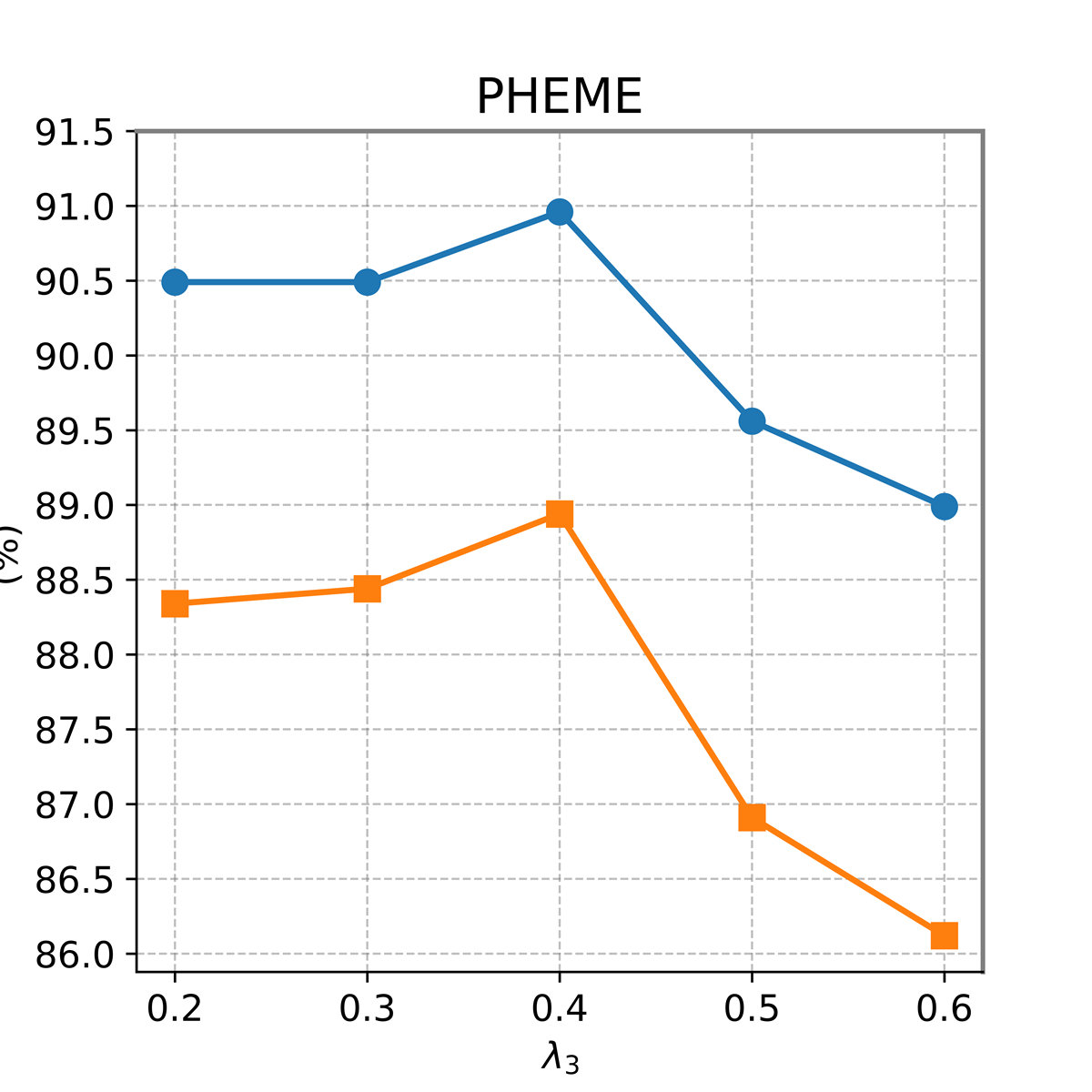}
		\label{subfig3_3}
	}
	\hfill
	\subfloat[]{
		\includegraphics[width=0.20\textwidth]{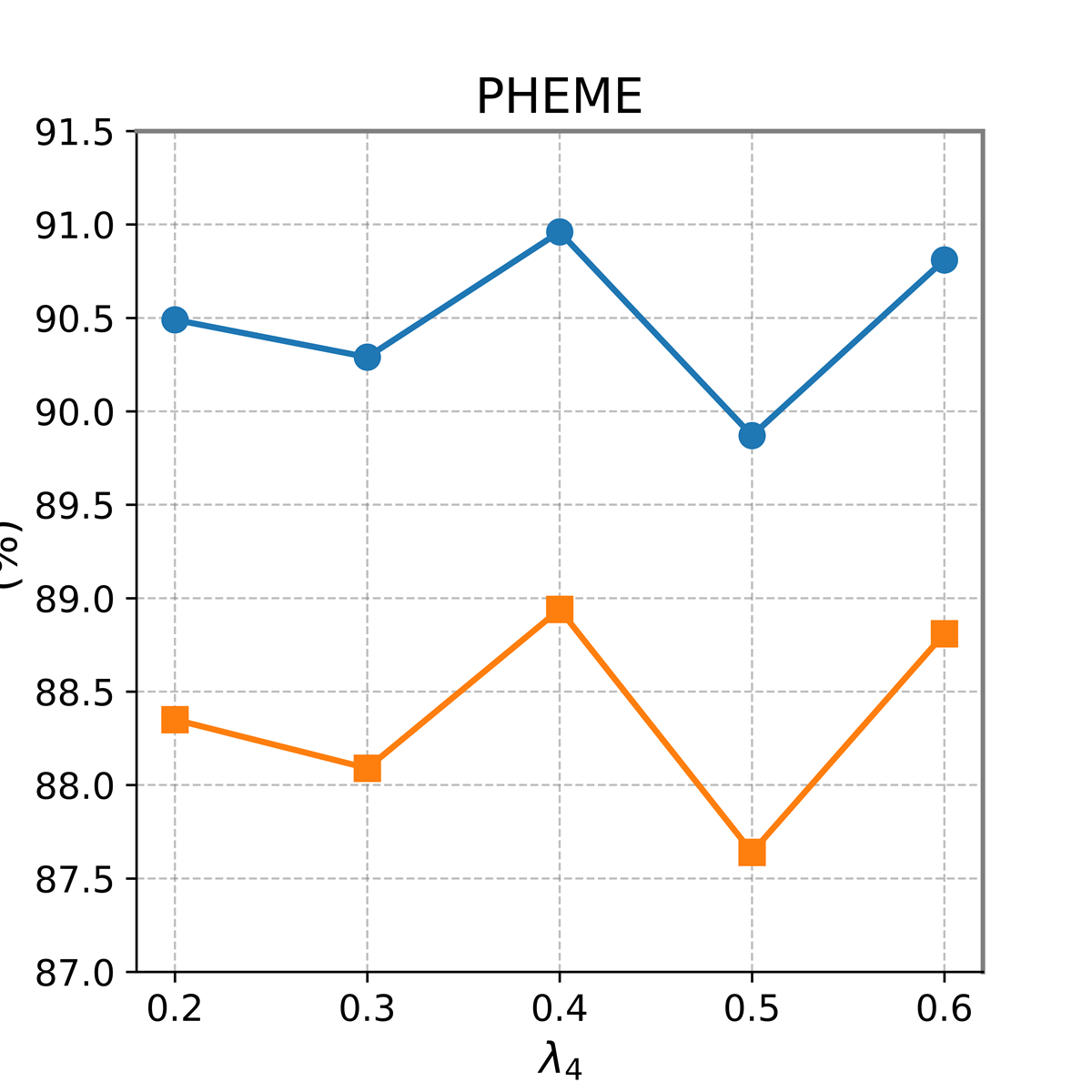}
		\label{subfig3_4}
	}
	\hfill
	\subfloat[]{
		\includegraphics[width=0.20\textwidth]{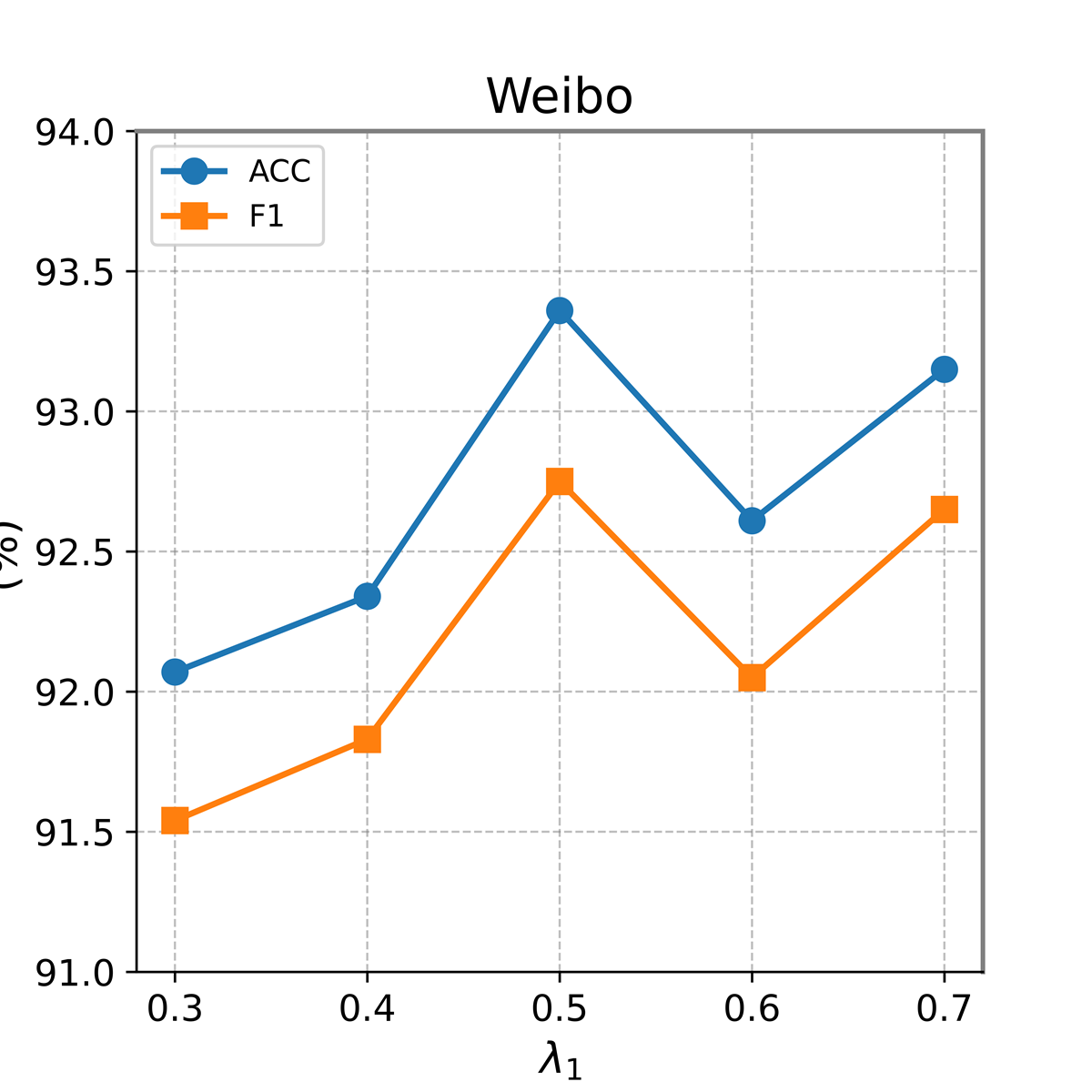}
		\label{subfig3_5}
	}
	\hfill
	\subfloat[]{
		\includegraphics[width=0.20\textwidth]{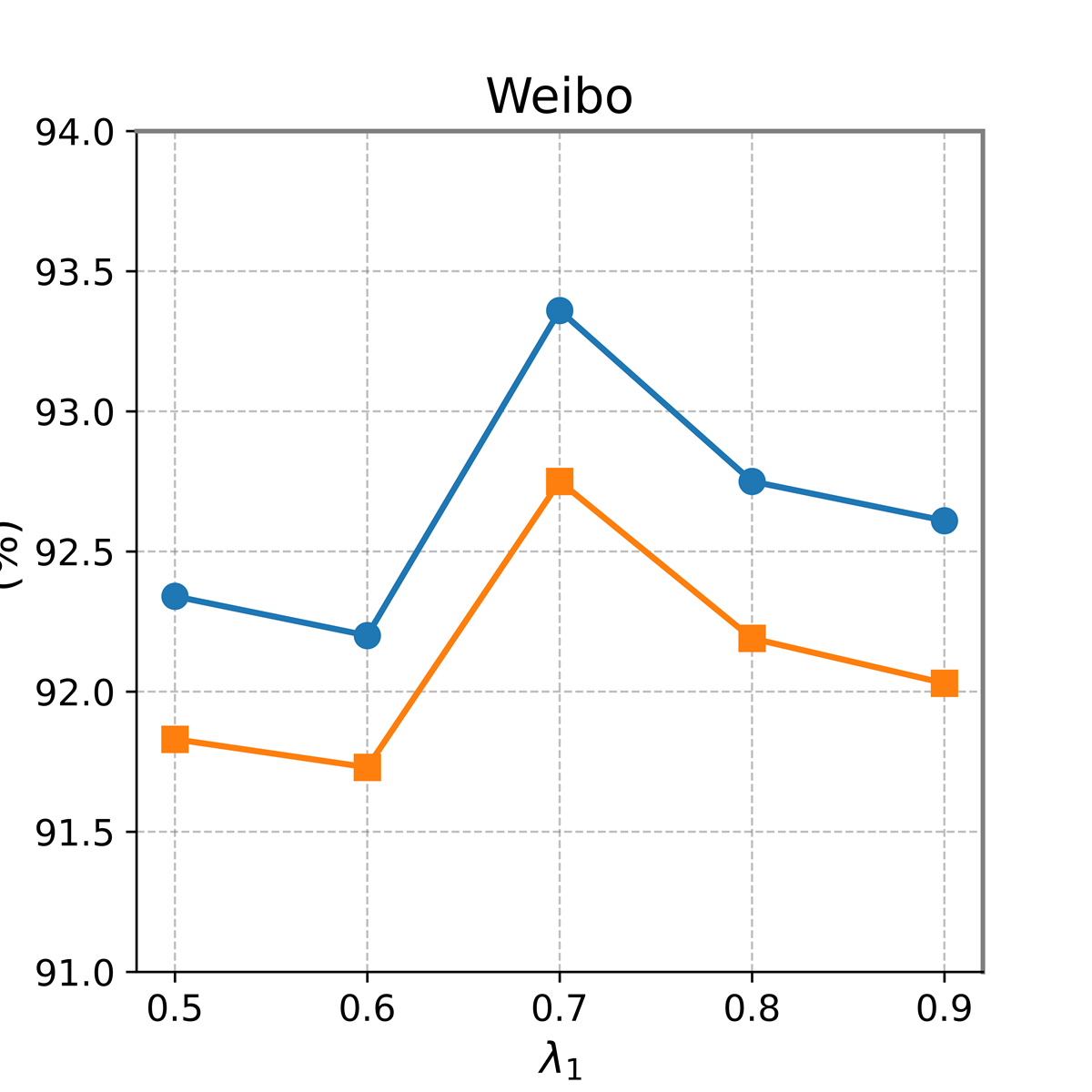}
		\label{subfig3_6}
	}
	\hfill
	\subfloat[]{
		\includegraphics[width=0.20\textwidth]{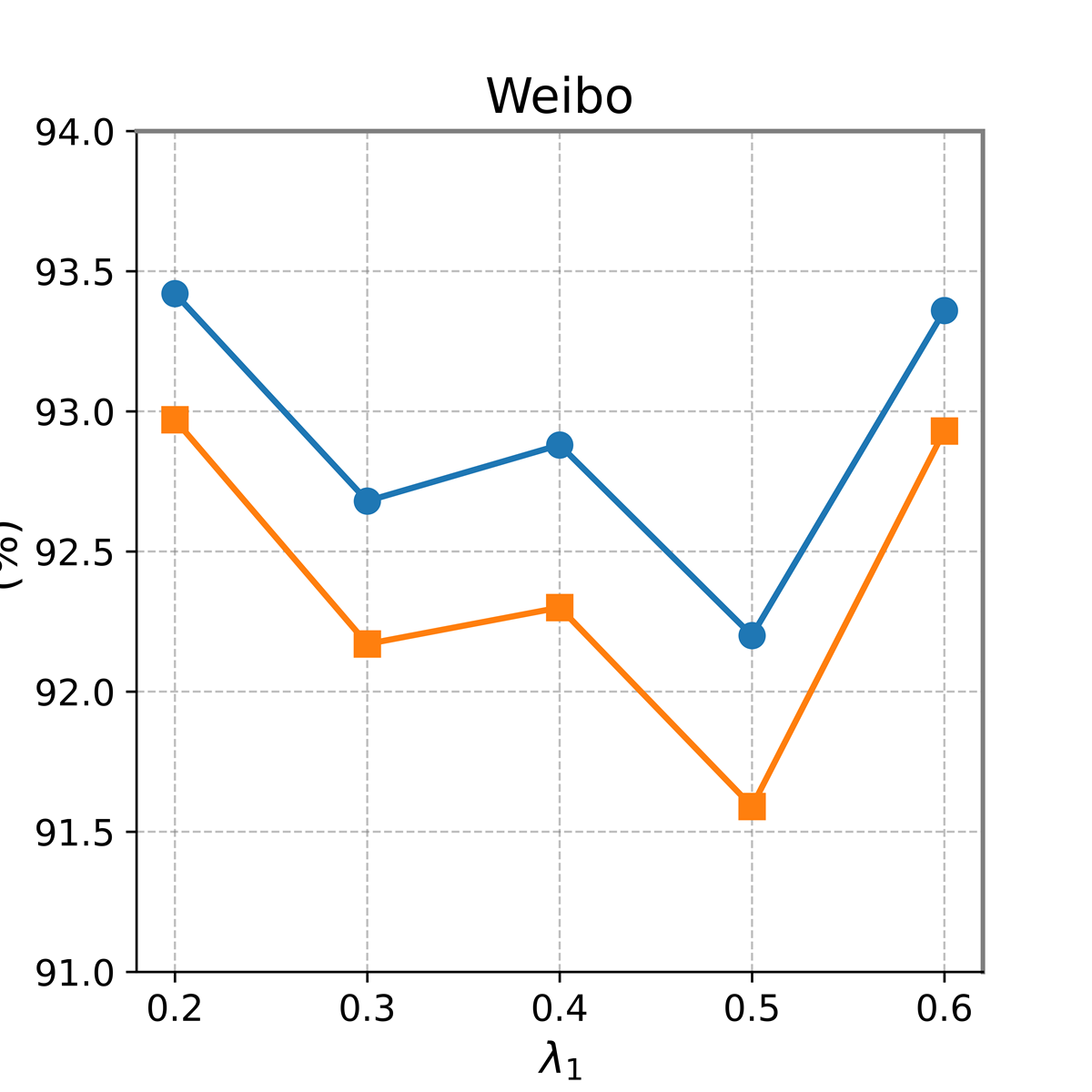}
		\label{subfig3_7}
	}
	\hfill
	\subfloat[]{
		\includegraphics[width=0.20\textwidth]{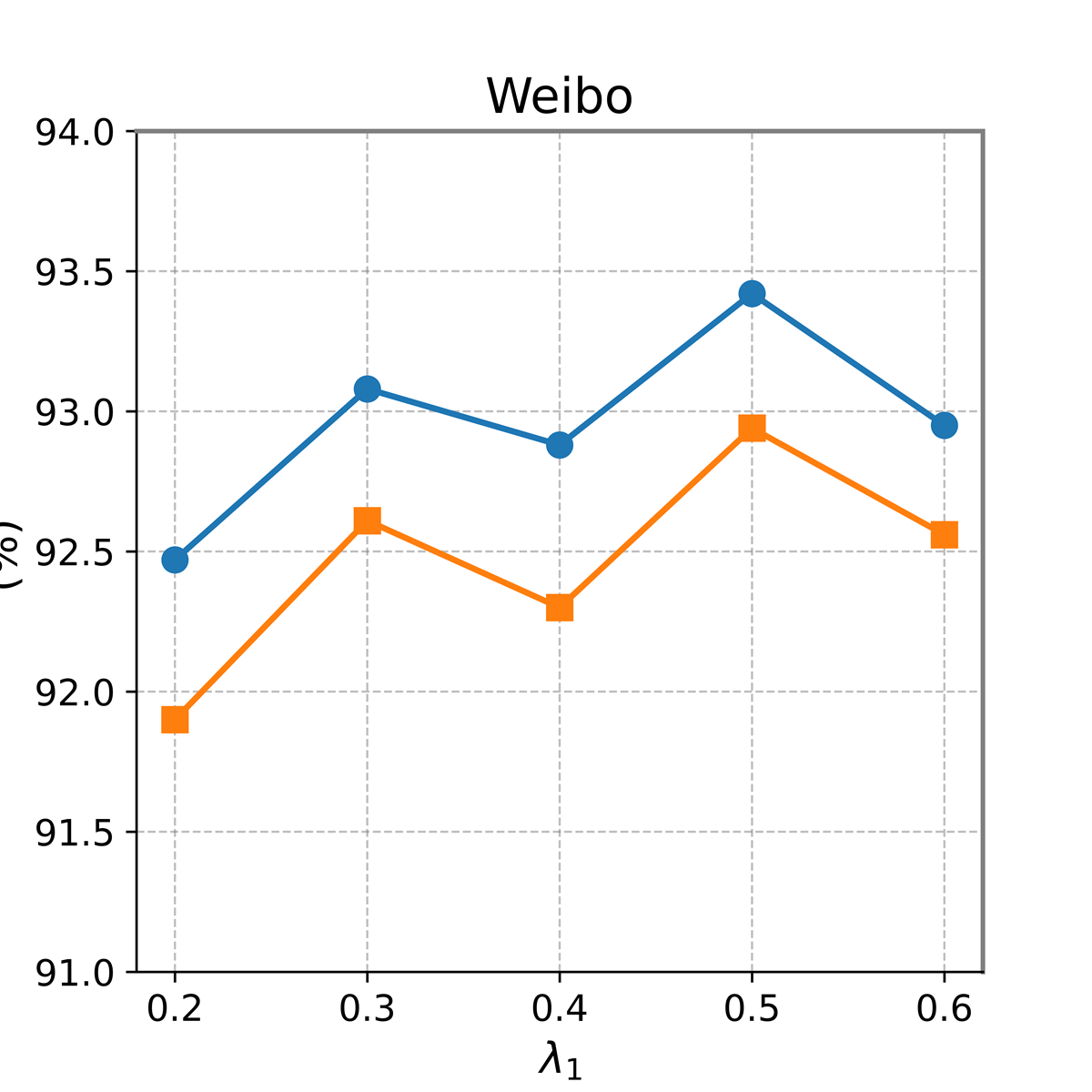}
		\label{subfig3_8}
	}
	\caption{Parametric analysis of the hyperparameter set \(\lambda\) on the PHEME and Weibo datasets, illustrating variations in accuracy (ACC) and F1-score (F1).}
	\label{fig3}
\end{figure*}

To assess the sensitivity of the proposed ISMAF to various hyperparameters, we perform a parametric analysis. Specifically, we vary the hyparameter set $\lambda = \{\lambda_1, \lambda_2, \lambda_3, \lambda_4\}$ of the overall loss function within the range of 0 to 1, with increments of 0.1, while keeping all other parameters fixed. Our preliminary experiments suggest that setting the parameter set around (0.3, 0.7, 0.4, 0.4) yields better model performance. To illustrate the effects of these variations, we visualized a representative subset of five distinct parameter sets in Fig.~\ref{fig3}.

The experimental results reveal that both accuracy and F1-score exhibit a highly consistent trend across varying hyperparameter settings, indicating that adjustments to \(\lambda\) primarily influence the model's classification performance in a consistent manner. Given that the classification loss is weighted at 1 in the overall loss function and classification is the model's primary task, excessively large hyperparameter values can lead to destabilizing the model's decision-making process, resulting in a decline in both ACC and F1-score. When the hyperparameter set is perturbed within a reasonable range around the selected values, ISMAF exhibits robust performance and consistently outperforms baseline methods on both datasets.

\section{Conclusion}\label{sec:con}
In response to the pressing need for robust rumor detection on social media, this paper introduces the Intrinsic-Social Modality Alignment and Fusion (ISMAF) framework, a novel approach that enhances multimodal rumor detection by effectively integrating textual, visual, and social context information. Unlike traditional models, which primarily focus on aligning intrinsic characteristics within news content, such as text and images, ISMAF addresses the complex interactions between intrinsic and social modalities. By categorizing text and image features as intrinsic modalities and social context features as the social modality, ISMAF provides a structured approach to cross-modality alignment, enabling a more comprehensive integration of diverse information sources.

Specifically, ISMAF employs a cross-modal consistency alignment strategy that addresses inconsistencies and enhances semantic coherence between intrinsic and contextual features. This alignment is complemented by a mutual learning mechanism that encourages collaborative refinement across modalities, enriching shared information and amplifying each modality’s individual contribution. Further advancing multimodal fusion, our adaptive fusion mechanism embedded within an encoder-decoder structure dynamically modulates the influence of each modality based on its relevance, achieving a flexible and efficient fusion compared to traditional methods.

Our extensive experiments on real-world English and Chinese datasets underscore the effectiveness of ISMAF, consistently outperforming state-of-the-art models in rumor detection tasks. Future work may extend ISMAF by incorporating additional data types, such as audio and video, to enrich feature diversity and enhance the framework’s adaptability. Moreover, integrating online learning capabilities could facilitate real-time detection, offering a timely response to misinformation and enhancing societal resilience against rumor proliferation.

\bibliographystyle{IEEEtran}
\bibliography{references}

\vfill

\end{document}